# Properties of a Molecular Cloud in NGC 185

L. M. Young

*Physics Department, New Mexico Institute of Mining and Technology, Socorro, NM 87801*

`lyoung@physics.nmt.edu`

## ABSTRACT

The dwarf elliptical galaxy NGC 185— the closest early-type galaxy with detectable molecular gas— offers a unique opportunity for high angular resolution studies of the interstellar medium in early-type galaxies. I present interferometric images (17 pc × 14 pc resolution) of CO emission from NGC 185. The majority of the molecular gas in NGC 185 is in one resolved cloud, closely associated with dust and an HI structure which may be a photodissociated envelope. The high spatial resolution also reveals a velocity gradient across the cloud, which allows a dynamical mass estimate independent of the $H_2$/CO conversion factor or the virial theorem. The linear sizes, velocity gradients, and dynamical masses of the molecular clouds in NGC 185 and NGC 205, a similar dwarf elliptical, are comparable to those of the largest Galactic giant molecular clouds (GMCs). If the dynamical masses of the clouds are assumed to be good estimates of their true masses, the inferred $H_2$/CO conversion factor in NGC 185 is at least eight times larger than a standard Galactic conversion factor, and the reasons for this difference are not yet understood. In any case, it seems clear that structures similar to Galactic GMCs can form in small elliptical or early type galaxies, even in the absence of spiral density waves.

*Subject headings:* galaxies: elliptical & lenticular, CD— galaxies: dwarf— galaxies:individual(NGC 185, NGC 205)— galaxies:ISM— Local Group— ISM:molecules

## 1. Introduction

Star formation is one of the fundamental aspects of galaxy evolution; in the most simplistic terms, the stars that form in a galaxy today help to determine what the galaxy will look like tomorrow. In this context, the molecular gas in elliptical or early-type galaxies is particularly interesting. The very name "early-type galaxy" suggests the paradigm that the ellipticals formed all their stars most of a Hubble time ago and then ceased, exhausting or ejecting all of their cold gas. But in recent years it has been found that many (if not most) ellipticals do contain dust and cold neutral gas; see Colbert et al. (2001), Goudfrooij et al. (1994), van Dokkum & Franx (1995), and Ferrari et al. (1999) (dust); Huchtmeier et al. (1995), Knapp et al. (1985), and Wardle & Knapp (1986) (HI); Rupen (1997) and Henkel & Wiklind (1997) (CO).

Thus, elliptical galaxies have the raw materials for star formation. Some authors (e.g. Kauffmann 1996) even suggest that if an elliptical galaxy accretes enough cold neutral gas, it could undergo a significant episode of star formation activity and transform itself into something that looks like a late-type galaxy. Clearly, a complete understanding of the evolution of elliptical galaxies must include a knowledge of their molecular gas, and specifically, where the gas came from and what it is doing.

The present paper focuses on the second of these issues— what the molecular gas is doing in elliptical galaxies. It is well known that molecular clouds in our own Galaxy and similar spirals form stars. But we have little information about the distribution and physical properties of the molecular gas in ellipticals. If the physical properties of the molecular gas in ellipticals (temperature, density, density structure, and others) are similar to those in Galactic GMCs, and if the molecular gas is in stable structures with lifetimes long enough to form stars, then star formation might proceed in ellipticals in a manner similar to spirals.

The main hindrance to studying the neutral ISM in ellipticals is that they tend to be rather far away, so the observed fluxes are weak and also the linear resolution is poor. The two major exceptions to this statement are two dwarf elliptical companions to M31, the galaxies NGC 185 and NGC 205. Their distances are less than 1 Mpc. They are the only early-type galaxies with a neutral interstellar medium that can be imaged at resolutions better than 100 pc, which allows us to resolve the individual giant molecular clouds and address the issues mentioned above.

It is true that these galaxies are not true ellipticals in some classification schemes; in general, dwarf ellipticals follow a different fundamental plane than giant ellipticals (Ferguson & Binggeli 1994), and many dwarf ellipticals have exponential light profiles rather than $R^{1/4}$. The dwarf ellipticals are not known to have significant amounts of hot gas, as many "true" ellipticals do. However, with respect to other factors which might influence the properties of their interstellar media, they can be regarded as ellipticals or as reasonable facsimiles thereof. They have mostly old stellar populations (Lee 1996; Davidge 1992; Martínez-Delgado et al. 1999), so the energetics of the interstellar medium are dominated by the evolved stars rather than the young massive stars. They are dynamically hot, supported by stellar velocity dispersions (Bender et al. 1991; Held et al. 1992), with no known signs of spiral density waves. Because NGC 185 and NGC 205 have these properties and others in common with "true" ellipticals, their interstellar medium offers important insights into the ISM of elliptical galaxies.

An interferometric map of the CO emission in NGC 205 (the first image which resolved an individual molecular cloud in an elliptical or dwarf elliptical galaxy) was presented by Young & Lo (1996). Various molecular line ratios in that galaxy were discussed by Young (2000) and Welch, Sage, & Mitchell (1998). The current paper discusses high resolution images of the molecular gas in NGC 185 and their implications for molecular cloud properties.

The first detection of CO emission from NGC 185 was published by Wiklind & Rydbeck (1986). It involved two pointings near the center of the galaxy, but the $33''$ beam was too large to say anything about the distribution of the molecular gas. Subsequent observations of the galaxy were all made with single dish telescopes at about $21''$ resolution. These include a map of the inner $40'' \times 80''$ of the galaxy, by Wiklind & Henkel (1993); eight pointings by Welch, Mitchell, & Yi (1996); and two pointings by Young & Lo (1997), hereafter YL97. The present paper gives an interferometric map of the CO emission in NGC 185, with the dual advantages of a larger field of view and higher spatial resolution than have been possible before. The high spatial resolution is particularly valuable— it resolves the molecular cloud, enables dynamical mass estimates of the cloud (without assuming a $H_2$/CO conversion factor), and reveals the relationships between molecular gas, atomic gas, and dust in this galaxy.

## 2. Observations and data reduction

The galaxy NGC 185 was observed in the $^{12}$CO J=1-0 line with the Berkeley-Illinois-Maryland Association (BIMA) millimeter interferometer in October 1996. At that time the array consisted of nine antennas arranged in a compact configuration, giving projected baselines from 2 to 50 k$\lambda$. A total of about 12 hours were obtained on source. Single sideband system temperatures (measured using the chopper wheel method and scaled to outside the atmosphere) were about 500-700 K for the seven SIS receivers and 700-900 K for the two Schottky receivers.

The phase and pointing center for these observations was 00h 38m 57.7s, $+48°\ 20'$ $12.0''$ (J2000 coordinates), approximately the optical center of the galaxy. NGC 185 is quite large in angular size— the optical isophotes are fit by an exponential with a scale length of $1.5'$ (Caldwell et al. 1992)— and the FWHM field of view of the BIMA interferometer is almost $2'$ at 115 GHz. A map of HI emission in NGC 185 shows that the atomic gas in NGC 185 is all located within $1.5'$ of the center (YL97). Since the CO extent is expected to be comparable to or smaller than the HI extent, and indeed the dust clouds in the galaxy are concentrated within about $30''$ of the galaxy center, only one pointing center was used for the CO observations. The correlator was set up to record data in the velocity range $-138$ to $-267$ km s$^{-1}$ (LSR velocities) at a resolution of 0.5 km s$^{-1}$.

Absolute flux calibration was based on observations of the planet Mars and double-checked with observations of the secondary calibrators 3C84 and BL Lac. The flux histories of those sources give flux calibrations which are consistent with Mars-derived values at a level of about 10%. The overall flux calibration is probably accurate to about 20%. The secondary calibrators 3C84 and BL Lac were used to track complex gain variations with time; they were observed about once every 45 minutes. Since the millimeter-wavelength continuum emission from NGC 185 is very weak and the CO line is also weak and relatively narrow, no attempt was made to correct the IF part of the bandpass shape.

Two spectral line cubes were made from the data, both with channels 4 km s$^{-1}$ wide and covering velocities from $-140$ km s$^{-1}$ to $-260$ km s$^{-1}$. All velocities in this paper are measured with respect to LSR. One cube was made with the so-called "natural" weighting of the uv data; it has an angular resolution of $5.5'' \times 4.6''$. The rms noise level in line-free regions of the cube is 0.070 Jy/beam (0.25 K). The other cube was made by applying a Gaussian tapering function to the uv data; it has a resolution of $9.8'' \times 8.2''$. Its rms noise level is 0.080 Jy/beam (0.091 K). Both cubes were lightly cleaned with the Hogbom clean algorithm. Image masks were created by smoothing and clipping the cleaned cubes, and integrated intensity maps were made by summing the data in unmasked regions of the cubes.

In this paper, the distance to NGC 185 is assumed to be $(m-M)_0 = 24.0 \pm 0.1$ mag, or $0.63 \pm 0.03$ Mpc. The value comes from Martínez-Delgado & Aparicio (1998), who derived $(m-M)_0 = 23.95 \pm 0.1$ mag from the tip of the red giant branch, and from Salaris & Cassisi (1998), who have compiled from the literature distance moduli in the range 24.06 to 24.15. At this distance, the linear resolutions of the two data cubes are $17 \times 14$ pc and $30 \times 25$ pc.

## 3. CO flux and distribution

Figure 1 shows the integrated intensity of molecular gas in NGC 185 superposed on an optical broadband ($B$) image and the atomic gas (HI) column density from YL97. Figure 2 shows individual channel maps for the CO emission.

The BIMA observations clearly show molecular gas in NGC 185 between velocities of $-188$ km s$^{-1}$ and $-200$ km s$^{-1}$. The CO emission is found in one cloud whose center coordinates are $00^h\ 38^m\ 56.6^s$, $+48°\ 20'\ 19''$ (J2000.0). The peak CO intensity is 0.31

Jy/beam = 1.1 K on the brightness temperature scale at $5''$ (15 pc) resolution. The highest CO integrated intensity is 3.3 Jy beam$^{-1}$ km s$^{-1}$= 11.9 K km s$^{-1}$, also at 15 pc resolution. The total flux of the cloud is 8.8 Jy km s$^{-1}$ with an uncertainty of at least 20% (see section 2). That number agrees very well with the IRAM 30m telescope observations of Young & Lo (1996), who found an integrated brightness temperature 2.1±0.2 K km s$^{-1}$ = 9.6±1.0 Jy km s$^{-1}$ in a $21''$ beam centered on the molecular cloud. The data of Wiklind & Rydbeck (1986) imply a flux of about 19 Jy km s$^{-1}$ in a larger beam, but without an error estimate, so it is difficult to judge whether that is inconsistent with the present results.

The CO distribution presented here agrees very well with the map of Wiklind & Henkel (1993), which was made with the IRAM 30m telescope and a Lucy deconvolution algorithm; both maps show the same prominent molecular cloud and very little (if any) CO elsewhere in the galaxy. The JCMT observations of Welch et al. (1996) show CO emission in three positions which all overlap the cloud in Figure 1; the velocities are consistent with those in the BIMA maps, so it is most probable that all sets of observations found the same gas. In short, there is no conclusive evidence that the interferometric images of the bright molecular cloud in NGC 185 have failed to detect a significant amount of molecular gas *in or around that cloud*. The interferometer data may lack the sensitivity to detect a smaller amount of gas elsewhere in the galaxy.

A Gaussian fit to the spectrum of the cloud (Figure 3) gives a center velocity of $-192.4\pm0.3$ km s$^{-1}$ and a FWHM of 18.3±0.6 km s$^{-1}$. Fitting the CO column density map with an elliptical Gaussian function gives the deconvolved FWHM of the molecular cloud to be $11\pm1'' \times 6.5\pm1''$ (34 pc × 20 pc); the same result is obtained from simply measuring the dimensions of the 50% contour of column density.

Adopting a H$_2$/CO conversion factor which is the same as a "standard" Galactic value, $X_{gal}= 3\times10^{20}$ cm$^{-2}$ (K km s$^{-1}$)$^{-1}$ (see Combes 1991), the peak H$_2$ column density is $3.6\times10^{21}$ cm$^{-2}$. At 0.63±0.03 Mpc, the H$_2$ mass of the cloud is $(4.1\pm0.9)\times10^4$ M$_\odot$.[1] Including helium increases the mass of the cloud by a factor of 1.36 (cf. Wilson 1995 and references therein), to $(5.6\pm1.2)\times10^4$ M$_\odot$. Welch et al. (1996) quote a mass which is 5–10 times higher than this, mainly because they assumed a H$_2$/CO conversion factor four times larger. In this paper, the mass of the molecular cloud (plus helium) determined from the CO luminosity and the conversion factor $X_{gal}$ is referred to as $M_{XCO}$. The appropriateness of this value $X_{gal}$ for external galaxies in general and for NGC 185 in particular is a matter of current debate; the mass of this molecular cloud is discussed in greater detail in sections 6, 7, and 9.

---

[1]All of the masses quoted in this paper have error estimates which fold in the distance uncertainty as well as other relevant measurement uncertainties.

## 4.    Comparison of CO to dust and HI

Optical broadband observations of NGC 185 reveal complex dust distributions near the center of the galaxy (Hodge 1963; Gallagher & Hunter 1981; Kim & Lee 1998). Hodge identified two relatively large, dark dust patches which he called Dust Cloud 1 (DC1) and Dust Cloud 2 (DC2); they can be seen most clearly in the beautiful images of Martínez-Delgado et al. (1999). Kim & Lee (1998) identify several other clouds which are smaller and/or more diffuse, but all of the known dust in NGC 185 is within 1′ of the center of the galaxy. Figure 4 shows that the molecular cloud I detected in NGC 185 is an extremely good match to the optical obscuration in Hodge's cloud DC1, in terms of position, size and shape.

Weak ($\leq 3\sigma$) detections of CO at positions on DC2 have been reported by YL97 and Welch et al (1996). The map presented here and the map of Wiklind & Henkel (1993) show no sign of CO emission near DC2. However, the reported detections in DC2 are factors of ~2 weaker than detections in DC1, so it is possible that molecular gas is present in DC2 and is too weak to be detected in the present dataset.

Welch et al. (1996) have asserted that a significant amount of the CO emission in NGC 185 comes from locations in the galaxy where dust obscuration is not seen. But the very close correspondence between the CO distribution and the dust obscuration in Figure 4, the agreement between the BIMA map and the IRAM 30m map of Wiklind & Henkel, and the fact that the strongest detections of Welch et al. all overlap Hodge's DC1 do not support this idea. It seems that all of the CO emission from NGC 185 can be attributed to regions with detectable dust obscuration.

The distribution and properties of the atomic gas (HI) in NGC 185 are discussed by YL97, who used the VLA to make images with resolutions of 21″ (64 pc) and 2.6 km s$^{-1}$. That work found $(1.09\pm0.06)\times10^5$ M$_\odot$ of HI in the galaxy,[2] or $(1.5\pm0.1)\times10^5$ M$_\odot$ including helium. All of the atomic gas is located within 2′ (370 pc) of the galaxy center. The peak HI column density coincides with galaxy center (within about 10″), and the HI has a cometary morphology; it falls off very quickly to the west, becoming undetectable about 30″ from the center, but trails off to distances of 2′ on the northeast (Figure 1).

The atomic gas is distributed in a handful of clumps which are easily seen in multiply-peaked HI profiles and in HI channel maps; the clumps show no global rotation pattern. They have masses around $10^4$ M$_\odot$, deconvolved radii around 50 pc, and velocity dispersions of 4–15 km s$^{-1}$. One HI clump of mass about $(9.5\pm1.9)\times10^3$ M$_\odot$ (including helium) and deconvolved FWHM 43″ × 32″ (130 pc × 100 pc) is coincident with the molecular cloud, within 5″ (15 pc) and 3 km s$^{-1}$ (Figure 4). This atomic clump may represent the photodissociated envelope around the molecular cloud. The peak HI column density in this clump is $1.7\times10^{20}$ cm$^{-2}$ (at a resolution of about 64 pc), and indeed the peak HI column density in NGC 185 as a whole is only about $3\times10^{20}$ cm$^{-2}$ (YL97). These values are significantly lower than the typical values of about $10^{21}$ cm$^{-2}$ which are usually thought to be required for the formation of molecular gas, at least in our own Galaxy; the difference may be caused by a much lower interstellar UV field in NGC 185 than in our own Galaxy, as discussed in greater detail by Young (2000).

---

[2]Corrected for the distance of 0.63 Mpc assumed here.

Welch et al. (1996) have argued that the atomic gas in NGC 185 is spatially anticorrelated with the molecular gas. Probably this impression is caused by the cometary morphology of the HI. But on small scales ($\sim$60 pc and smaller), it seems that the atomic and molecular phases are indeed associated with each other, just as one would normally expect a molecular cloud to be associated with some photodissociated atomic gas.

## 5. Previous observations of NGC 205

For purposes of comparison, observations of CO emission in NGC 205 (first presented by Young & Lo 1996) were re-imaged at a velocity resolution of 2 km s$^{-1}$ and angular resolutions of $9.55''\times5.0$ '' and $17.0''\times12.8$ ''. A distance of 0.85 Mpc is assumed for NGC 205, and the uncertainty in this distance is probably at least 0.1 Mpc (see Salaris & Cassisi [1998] and Ferrarese et al. [2000] and references therein). At this distance, the linear resolutions of the images are 39 pc $\times$21 pc and 70 pc $\times$53 pc. CO emission in NGC 205 was analyzed in the same manner as for NGC 185; the CO flux detected in NGC 205 is 10 Jy km s$^{-1}$ $\pm$20%, in good agreement with the 11 Jy km s$^{-1}$ $\pm$14% from the single dish observations of Young (2000). The H$_2$ mass calculated from a H$_2$/CO conversion factor is therefore $(8.5\pm2.6)\times10^4$ M$_\odot$, not including helium, or $M_{XCO}= (1.16\pm0.36)\times10^5$ M$_\odot$ including helium. A Gaussian fit to the integrated CO spectrum of the cloud gives a FWHM 7.3$\pm$0.3 km s$^{-1}$, in reasonable agreement with the values obtained at much higher signal-to-noise ratio from the IRAM 30m telescope (Young 2000). An elliptical Gaussian fit to the CO column density map gives the deconvolved FWHM of the cloud to be 15$\pm$1$''$ $\times$ 6.6$\pm$1$''$ (62 pc $\times$ 27 pc). The molecular cloud is also associated with an HI cloud of mass $(7.3\pm0.7)\times10^4$ M$_\odot$ and with dust obscuration (see Young & Lo [1996]; Young [2000]).

## 6. Velocity gradients and rotational mass estimates

A quantitative understanding of star formation and the ISM in NGC 185 and NGC 205 clearly hinges on the mass estimates for the molecular clouds. But the $H_2$/CO conversion factor which was used in section 3 might not be applicable to the dwarf ellipticals. A dynamical mass estimate would provide a valuable check on the physical properties of the gas in NGC 185 and NGC 205 as well as on the general use of $H_2$/CO conversion factors in different environments.

A virial calculation is a very popular dynamical mass estimate, and virial masses for the clouds in NGC 185 and NGC 205 will be discussed in section 7. However, as the discussion of Maloney (1990) shows, it's not clear whether virial equilibrium is a good assumption to make for the clouds in our own Galaxy. And if that is true in our own Galaxy, it is even *more* true in the dwarf ellipticals, where even less is known about the gas properties, interstellar pressure, magnetic fields, and cloud lifetimes. Fortunately, the high linear resolution provided by the BIMA array allows us to make a dynamical mass estimate for a molecular cloud *without using either a conversion factor or an assumption about virial equilibrium*. Specifically, the molecular clouds DC1 in NGC 185 and DC11 in NGC 205 show significant velocity gradients; if these are interpreted as rotation, they give an estimate of the gravitating mass of the cloud in the same way that is usually done for rotating galaxies.

The channel maps in figures 2 and 5 clearly show a velocity gradient in the north-south direction for NGC 185 and in the east-west direction for NGC 205, roughly aligned with the cloud's major axis in each case. The velocity gradients were measured from position-velocity (p-v) plots (Figures 6 and 7). The position angles for these measurements were chosen by eye, based a combination of isovelocity contours on the intensity-weighted mean velocity maps and the cloud's major axis in the total column density maps. The kinematic and morphological major axes agree within 20° for DC11 in NGC 205. They are offset by about 40° in NGC 185, but both position angles give similar mass estimates (the kinematic position angle gives a larger velocity gradient but a smaller cloud radius). The velocity gradient, extent and velocity width of each rotating region were measured on the approximate midline of the cloud in the p-v diagram and the contour level at 10% of peak intensity (see figures 6 and 7). Uncertainties were roughly estimated from the range of several different reasonable-looking values. Of course these procedures give only crude measurements, but since the clouds are only a few resolution elements in diameter, more elaborate rotation curve-fitting procedures would probably not be helpful or justified.

The velocity gradients, cloud radii, and projected rotation velocities are given in Table 1. The implied gravitating mass inside the radius $R$ is calculated from $M_{rot} = V^2 R/G$, or $M_{rot} = (233\,M_\odot)V^2 R$ with $V$ in km s$^{-1}$ and $R$ in pc. (As usual, $M_{rot}$ is a lower limit because of the unknown inclination angle between the rotation axis and our line of sight.)

Like the clouds in NGC 185 and NGC 205, Galactic GMCs often show significant velocity gradients along their longest direction (Blitz 1993; Phillips 1999). In fact, the molecular clouds in NGC 185 and NGC 205 have diameters, velocity gradients, specific angular momenta, and (therefore) dynamical masses which are similar to those of the largest Galactic GMCs (most notably L1641 and the Rosette Molecular complex; see Phillips [1999]). The average densities of the clouds in NGC 185 and NGC 205 are also similar to the average densities of Galactic GMCs; for NGC 205, values of 10–190 $H_2$ molecules $cm^{-3}$ are found, depending on what values of mass and radius are assumed (see Table 1). For the cloud in NGC 185, average densities fall in the range 25–2500 $cm^{-3}$.

For Galactic GMCs there is usually a great deal of vacillation about whether a velocity gradient is caused by rotation or by something else— gas outflows, differential acceleration driven by an asymmetric distribution of young massive stars, and so on. As a result, rotational masses are rarely calculated for Galactic GMCs. But Blitz (1993) and Phillips (1999) offer some arguments that velocity gradients in Galactic GMCs are usually due to rotation, at least in the large clouds. Phillips (1999) shows that such rotation is usually dynamically important for clouds larger than a few pc, like the ones in NGC 185 and NGC 205.

One might hope that the position-velocity diagrams could be used to constrain the mass distribution within the molecular clouds, in the same way that rotation curves of galaxies constrain their mass models. However, the appearance of solid-body rotation is determined more by the poor linear resolution and by the fact that the CO emission is optically thick than by the actual mass distribution within the cloud.

## 7. Virial mass estimates

Defining linewidths and especially sizes of irregular objects like molecular clouds is not straightforward, and as a result virial mass estimates for molecular clouds have been calculated in many different ways. In this paper I have followed the methods of Wilson and Scoville (Wilson & Scoville 1990, 1992; Wilson 1995), who studied giant molecular clouds in the Local Group galaxies M33, NGC 6822, and IC 10. Since these galaxies are at approximately the same distance as NGC 185 and NGC 205, and were observed with approximately the same sensitivity and resolution, the cloud diameters, linewidths, and virial masses will be directly comparable.

As usual, Wilson & Scoville (1990, 1992) ignore magnetic field support and external pressure confinement. one The clouds are assumed to have density profiles which drop as

$r^{-1}$. Under these conditions, Wilson & Scoville calculate $M_{vir}$ from

$$M_{vir} = (99 \ M_\odot) \ \Delta v^2 D$$

, where the linewidth $\Delta v$ is in km s$^{-1}$ and the cloud's diameter $D$ is in pc. (Note that MacLaren et al. [1988] give the constant as 95 M$_\odot$ instead of 99 M$_\odot$.) Table 1 gives $D$, $\Delta v$, and $M_{vir}$ for the clouds in NGC 185 and NGC 205.

The diameters of the molecular clouds in NGC 185 and NGC 205 are similar to those in the other Local Group galaxies, and the same is true of the linewidth of the cloud DC11 in NGC 205. However, the CO linewidth in NGC 185 is much larger than that of any other cloud in the Local Group sample— about 40% larger than its closest competitor. As a result, the cloud in NGC 185 sits far above the size-linewidth relation determined by Wilson & Scoville (1990) and its virial mass is about a factor of two larger than the largest other cloud in the comparison sample.

Because the dynamical masses of the molecular clouds depend sensitively on the linewidth, and the linewidth in NGC 185 is surprisingly large, that measurement deserves extra scrutiny. The signal-to-noise ratio in the BIMA spectra is not particularly high, either for NGC 185 or for NGC 205, but the width of the line in NGC 205 is confirmed by much higher quality spectra from the IRAM 30m telescope (Young 2000). Thus, the signal-to-noise ratio in the BIMA spectra seems to be high enough to make a reliable determination of line width. (The formal uncertainty estimate for the width, 0.6 km s$^{-1}$ for NGC 185, is probably lower than the true uncertainty which would include systematic effects such as summing over a slightly different area on the sky.) The large CO linewidth in NGC 185 is also corroborated by previous single-dish CO observations of YL97, which found linewidths of about 18 km s$^{-1}$ in NGC 185 DC1, and by Welch et al. (1996), who found an even larger value. The HI cloud which is associated with DC1 also has a linewidth around 20 km s$^{-1}$ (YL97). Finally, the individual channel maps (Figure 2) provide clear evidence of a large linewidth in NGC 185, even without line fits.

## 8. Is the cloud DC1 in NGC 185 in dynamical equilibrium?

As mentioned in the introduction, the amount of star formation that can happen in these molecular clouds depends on whether they are stable entities with dynamical lifetimes greater than the time necessary for the star formation process. Calculations of $M_{rot}$ and $M_{vir}$ *assume* that the clouds are in dynamical equilibrium; but since so little is known about molecular clouds in ellipticals, this assumption is worth checking, and particularly so for DC1 in NGC 185, where the large linewidth implies large dynamical masses.

The stellar population of NGC 185 was investigated most recently by Martínez-Delgado, Aparicio, & Gallart (1999), who find that there was a small amount of star formation in the center of that galaxy as recently as $10^8$ years ago (see figure 4). Very little, if any, star formation has occurred since that time. Because of the lack of current star formation activity it seems unlikely that the linewidth of DC1 is inflated by the kinetic energy input of young massive stars.

However, the cloud might be passing close enough to the center of the galaxy to be tidally disrupted. If a small cloud of mass $m$ and radius $r$ orbits at a distance $R$ from a mass $M$, and $r \ll R$, then $m > 2M(r/R)^3$ is required to prevent tidal disruption of $m$ (e.g. von Hoerner 1957; Shu 1990). The molecular cloud DC1 has a radius $r \sim 20$ pc; its projected distance $R$ from the center of the galaxy is about $13''$ or 40 pc. The mass $M$ of the galaxy interior to the projected radius $R = 13''$ can be estimated as follows. Caldwell et al. (1992), reanalyzing the data of Kent (1987), find NGC 185 well fit by an exponential light distribution with a scale length of $1.5' = 275$ pc and a total magnitude $M_V = -15.3$. For that scale length, about 1% of that light is within a radius of $13''$. Peterson & Caldwell (1993) find a mass-to-light ratio of 5.3 in $V$, which (if constant throughout the galaxy) implies a mass of $6 \times 10^8$ $M_\odot$ for the entire galaxy and $6 \times 10^6$ $M_\odot$ interior to $13''$.

For the cloud DC1 to be stable against tidal disruption according to the formula given above, it must have a mass greater than about $1.5 \times 10^6$ $M_\odot$. A slightly better calculation, made without assuming $r \ll R$, gives a somewhat smaller limit of $8 \times 10^5$ $M_\odot$. This value is comparable to the estimated $M_{vir}$ and larger than $M_{rot}$ and $M_{XCO}$ for the cloud DC1, implying that indeed the cloud may be vulnerable to tidal disruption. The large CO linewidth in NGC 185 could therefore be a symptom of a cloud which is being disrupted, rather than an indicator of a large mass. Of course, the true distance from the cloud to the center of the galaxy is probably larger than the projected distance of $13'' = 40$ pc, so the cloud could be stable for that reason.

At present the dynamical status of the cloud DC1 in NGC 185— whether it is in an orbit where it will be stable— cannot be decided because of the unknown geometry; however, it is clear that disruption in the near future is a possibility for this cloud. In contrast, the cloud DC11 is considerably farther away from the center of its galaxy, NGC 205 ($53'' = 220$ pc in projection). That cloud is much less likely to suffer tidal disruption than DC1 in NGC 185, provided it stays out of radial orbits.

## 9. H$_2$/CO conversion factor

If the molecular clouds are assumed to be in virial equilibrium or to be supported by rotation (i.e. if the linewidth traces the gravitating mass of the cloud, not the process of tidal disruption), then the dynamical mass ($M_{vir}$ or $M_{rot}$) is taken to be the true mass of each cloud. One can then use the observed CO luminosity and $M_{XCO}$ to infer $X_{vir}$ and $X_{rot}$, CO-to-H$_2$ conversion factors which are appropriate for NGC 185 and NGC 205 (Table 1):

$$\frac{X_{rot}}{X_{gal}} = \frac{M_{rot}}{M_{XCO}}; \frac{X_{vir}}{X_{gal}} = \frac{M_{vir}}{M_{XCO}}.$$

The H$_2$/CO conversion factors which are inferred for NGC 205 are similar to the adopted value of $X_{gal}$, but the inferred values for NGC 185 are about ten times larger than $X_{gal}$; this cloud is significantly underluminous in CO, compared to its Galactic counterparts, if the mass estimates are accurate. This large discrepancy cannot be convincingly explained at the present time. Many authors assert that the H$_2$/CO conversion factor should increase as the metallicity of a galaxy decreases (e.g. Wilson 1995, Israel 1997, Mihos et al. 1999). Planetary nebulae in NGC 185 and 205 have oxygen abundances about 0.2 solar and 0.5 solar, respectively (Richer & McCall 1995); thus, if these galaxies followed the trend with metallicity of the other Local Group galaxies in Wilson (1995), their conversion factors would be $X_{185}/X_{gal}$= 2.5 and $X_{205}/X_{gal}$= 1.6. This metallicity effect may adequately explain the properties of the cloud in NGC 205, but it is clearly not large enough, by itself, to explain the inferred conversion factors in NGC 185. (If the cloud in NGC 185 is not in dynamical equilibrium, of course, this conversion factor problem disappears.)

Metallicity is not the only property which may affect the conversion factor (e.g. Walter 2001); for example, the conversion factor is also expected to change with the strength of the local UV field. NGC 185 and NGC 205 have internal UV fields which are estimated at between 0.01 and 0.1 times the "standard solar neighborhood" value (Welch et al. 1996, 1998). But exactly how the conversion factor changes with the UV field is not clearly established. Israel (1997), Lequeux et al. (1994), and others have argued that a high UV field in the SMC preferentially destroys CO instead of H$_2$ and raises the conversion factor. On the other hand, Weiss et al. (2001) argue that a high UV field increases the temperature of the molecular gas in the center of M82. The CO luminosity is thereby increased, and the conversion factor is lowered. Cases of molecular clouds bathed in very low UV fields are rarely discussed in the literature. Thus it is difficult to predict whether the effect of the low UV field in NGC 185 would be in agreement with observations.

The large conversion factor in NGC 185 could conceivably be caused by a different density structure than the other clouds in the Local Group sample; specifically, homogeneous clouds (ones with a small clump-interclump density ratio) can have a conversion factor which is ten times larger than their clumpy counterparts (Mihos et al. 1999). This possibility could be checked with observations of molecular isotopic line ratios, which would give another independent density estimate.

## 10. Implications for the Formation of Molecular Clouds

A global understanding of star formation in galaxies depends on the evolution of the interstellar medium and the life cycle of molecular clouds, i.e. how the clouds are formed and destroyed. Heyer & Terebey (1998) argue that the molecular clouds which now exist in the spiral arms of our Galaxy were created out of atomic gas by the shocks associated with the arms. Other authors (see Heyer & Terebey for references) have argued that the molecular gas is merely assembled, not created, by the density waves. In either case, it is clear that the density waves are assumed to play a critical role in the origin of GMCs in our Galaxy.

Contrasting perspectives on the importance of spiral density waves are offered by the flocculent galaxies, disk galaxies in which the grand-design spiral density waves are very weak (if present at all; e.g. Elmegreen et al. 1999, Thornley 1996). The flocculent galaxies NGC 5055 and NGC 4414 do show large, Giant Molecular Association (GMA)-type structures similar to those in our own Galaxy (Thornley & Mundy 1997a, 1997b). This result is particularly interesting for the case of NGC 4414, in which the GMAs have formed *without* strong, long-lived kinematic arms; apparently local instabilities or other mechanisms can form molecular clouds, and spiral density waves are not required.

The present observations of CO in NGC 185 and NGC 205 may strengthen this idea, if the gas has an internal origin. We know of no spiral density waves in the two dwarf ellipticals; indeed, these galaxies are dynamically hot, supported by stellar velocity dispersions rather than by rotation (Bender et al. 1991; Held et al. 1992; Peterson & Caldwell 1993). It is possible, of course, that the GMCs in NGC 185 and NGC 205 were acquired from some larger disk galaxy through gravitational interaction (e.g. Knapp et al. 1985), and that they were originally formed in density waves. On the other hand, Oosterloo et al. (1999) suggest that the neutral gas in the smaller early-type galaxies may have an internal origin, which would be consistent with the distribution and kinematics of the HI in NGC 185 (YL97).

## 11. Summary

High resolution interferometric observations of CO emission in NGC 185 show that the detected molecular gas is concentrated in one GMC-size structure. It has a CO flux of 8.8 Jy km s$^{-1}$ $\pm20\%$, and the molecular mass is $(5.6\pm1.2)\times10^4$ M$_\odot$, including helium, if one assumes a H$_2$/CO conversion factor of $3\times10^{20}$ cm$^{-2}$ (K km s$^{-1}$)$^{-1}$. The cloud has deconvolved FWHM diameters 34 pc $\times$ 20 pc; it is exceedingly well matched in size and shape to the darkest patch of optical dust obscuration in the galaxy. The molecular gas is most likely associated with a clump of atomic gas of mass $(9.5\pm1.9)\times10^3$ M$_\odot$, including helium. The peak HI column density of this clump is $1.7\times10^{20}$ cm$^{-2}$, much lower than the HI column density which is usually thought to be required for the formation of molecular gas.

The high linear resolution (17 pc $\times$ 14 pc) reveals a velocity gradient, suggestive of solid body rotation, roughly aligned with the major axis of the cloud. In this respect the CO cloud in NGC 185 is similar to the cloud DC11 in NGC 205. The velocity gradients are estimated to be $0.47\pm0.09$ km s$^{-1}$ pc$^{-1}$ and $0.16\pm0.03$ km s$^{-1}$ pc$^{-1}$ for the clouds in NGC 185 and NGC 205 respectively; the gradients extend over 40 pc and 70 pc, giving gravitating masses of $(4.2\pm1.0)\times10^5$ M$_\odot$ and $(2.6\pm0.6)\times10^5$ M$_\odot$ respectively. Virial masses are similar to those rotational masses but may not be relevant because of the large rotational component to the line width.

H$_2$/CO conversion factors derived from those dynamical masses are at least $7.5\pm2.4$ (for NGC 185) and $2.4\pm0.7$ (for NGC 205) times larger than the "standard" Galactic conversion factor mentioned above. The conversion factor for NGC 205 can be explained by that galaxy's metallicity and the proposed variation of conversion factor with metallicity; however, the conversion factor in NGC 185 is too large to be explained by metallicity effects alone. If the clouds are in dynamical equilibrium, the cloud in NGC 185 is significantly underluminous compared to its Local Group counterparts. However, its projected distance to the center of the galaxy is small enough to suggest that it may be suffering tidal disruption, leading to a large linewidth.

The major result of the present paper, especially in combination with the work of Young (2000) on molecular line ratios in NGC 205, is that the molecular gas in these two dwarf elliptical or early-type galaxies is found in structures which are very much like Galactic GMCs. The major known differences between these clouds and Galactic GMCs are (1) the very low HI column densities which are associated with molecular gas, and (2) the large inferred H$_2$/CO conversion factor in NGC 185, which admittedly could be overestimated if the cloud is not in dynamical equilibrium. To the extent that the physical properties of the molecular clouds in the dwarf ellipticals are similar to the properties of Galactic GMCs, there is no reason not to suppose that star formation may proceed in the dwarf ellipticals in a manner similar to star formation in Galactic clouds.

Thanks to Liese van Zee for providing the optical images of NGC 185 and to Michael Rupen, Christine Wilson, and Rick Forster for helpful discussions. This material is based upon work supported by the National Science Foundation under grant AST-0074709.

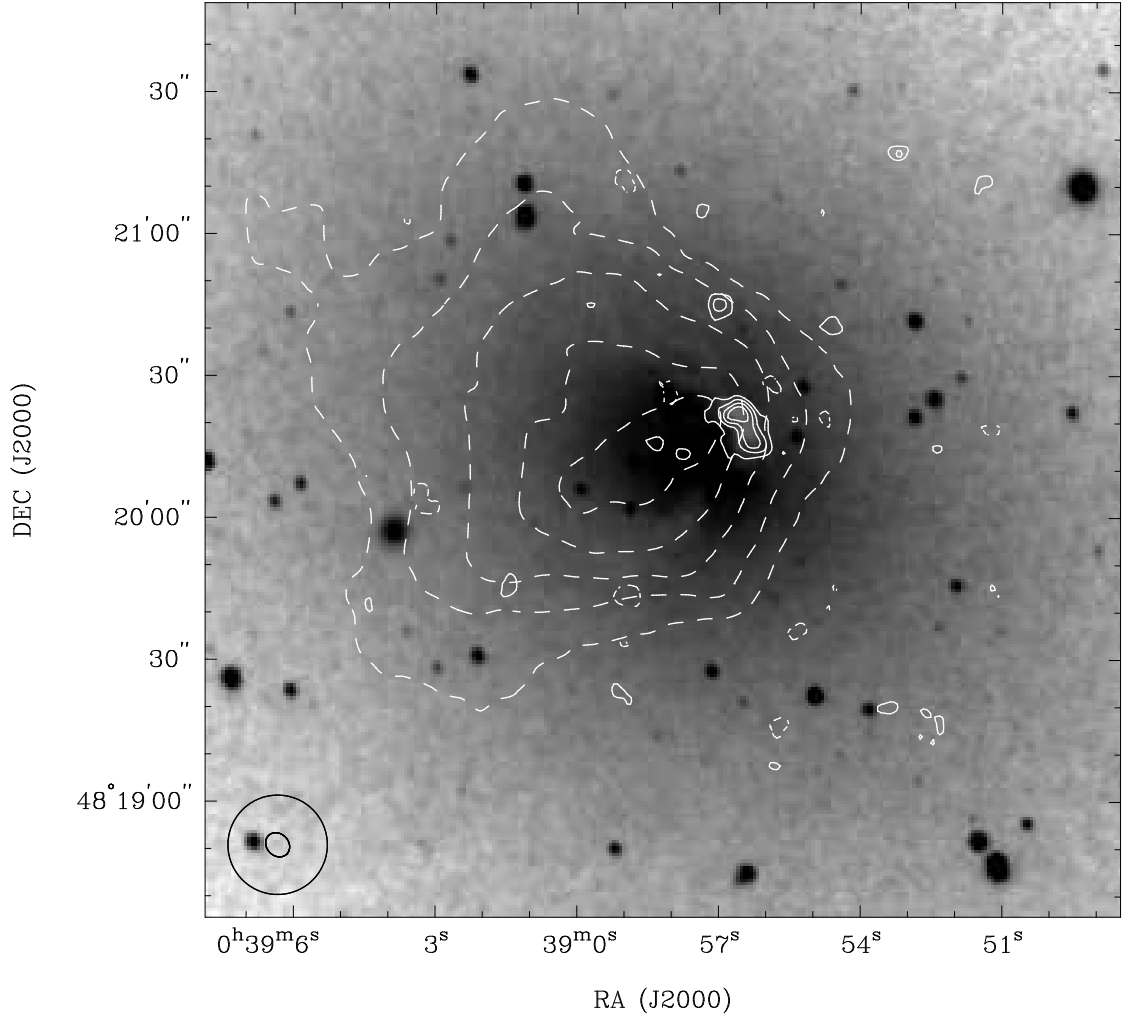

Fig. 1.— CO and HI on an optical image of NGC 185. The greyscale is a B image of NGC 185 from the KPNO 0.9m telescope, courtesy of L. van Zee. The dashed contours show the total HI column density, with contour levels at 10, 30, 50, 70, and 90 percent of the peak ($3 \times 10^{20}$ cm$^{-2}$). The solid contours show the CO integrated intensity, with contour levels at $-20$, 20, 40, 60, and 80 percent of the peak (12 K km s$^{-1}$). HI and CO beam sizes are indicated in the bottom right corner.

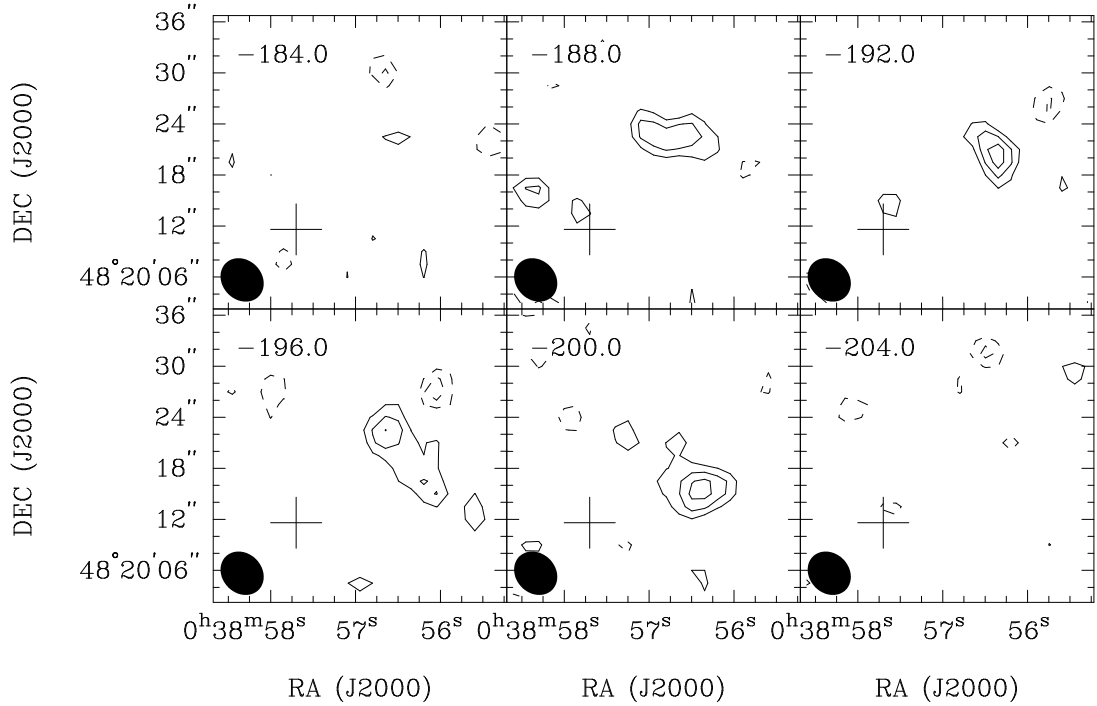

Fig. 2.— Channel maps of the CO emission in NGC 185. The contour levels are −3, −2, 2, 3, 4 times the rms noise level (about 0.25 K). Each channel is 4 km s$^{-1}$ wide and labelled with its center velocity (LSR). The beam size (5.5″× 4.6″; this is the high resolution cube) is indicated with a filled ellipse. The cross marks the approximate optical center of the galaxy, as given by the NASA Extragalactic Database.

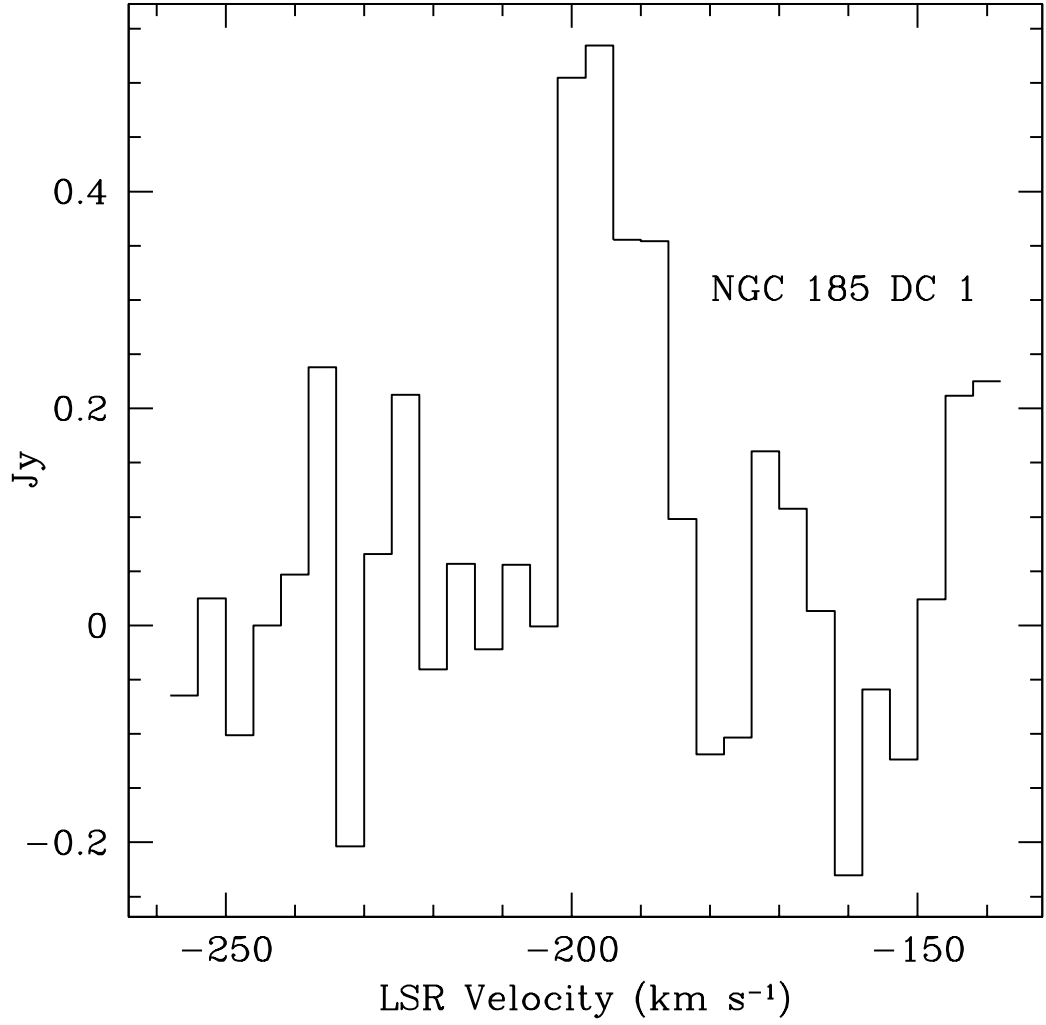

Fig. 3.— Spectrum of CO emission from NGC 185 DC 1. The spectrum is a simple sum over most of the cloud area.

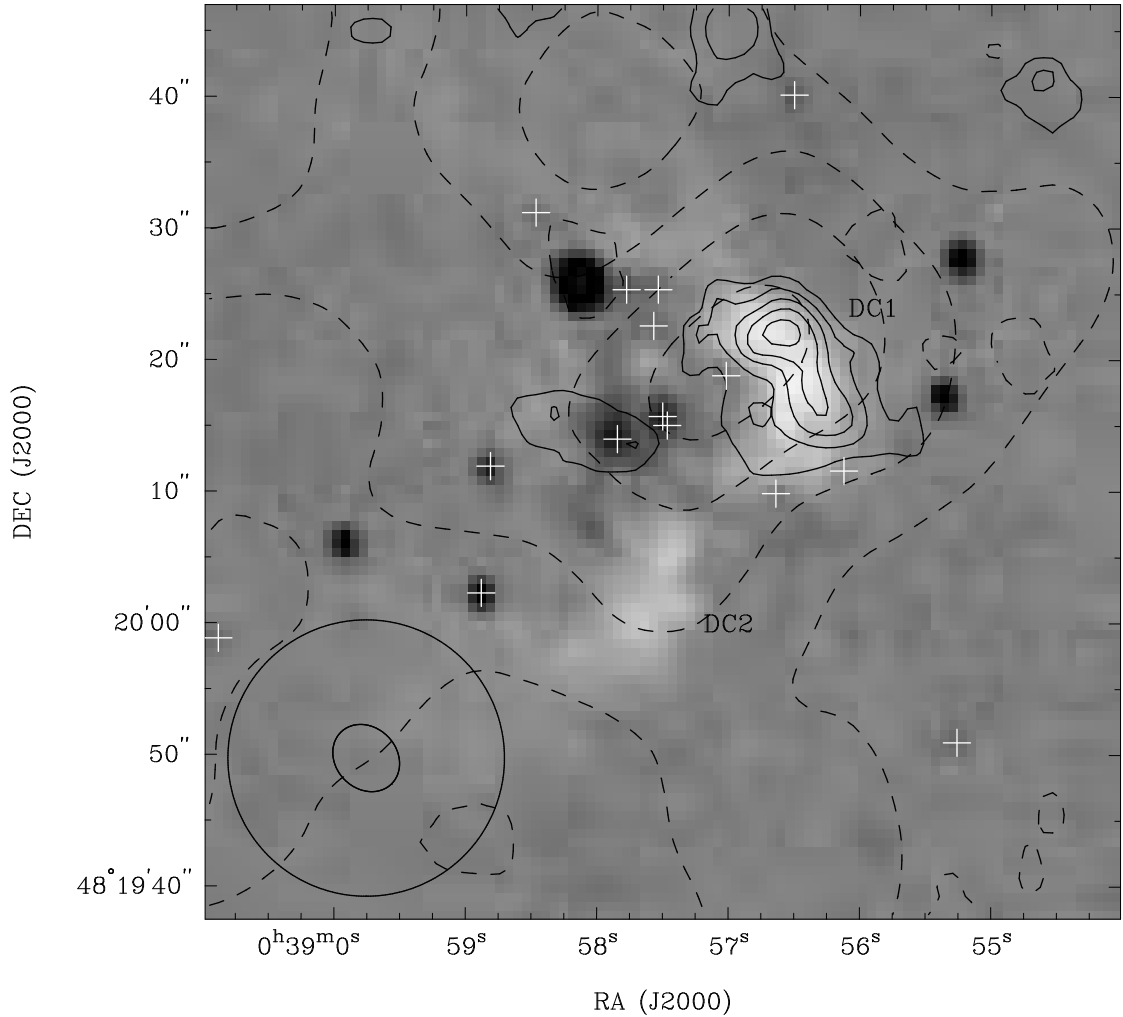

Fig. 4.— CO, HI, and dust in NGC 185. The greyscale is a B image of NGC 185 (Figure 1) from which a smooth elliptical model has been subtracted. Dust clouds are the white patches. Solid contours show the integrated intensity of CO emission, with contour levels at $-10$, 10, 30, 50, 70, and 90 percent of the peak (12 K km s$^{-1}$). Dashed contours show the HI column density, this time integrated only over the velocity range in which the CO emission is found. Contour levels are 30, 50, 70, and 90 percent of $1.7 \times 10^{20}$ cm$^{-2}$. HI and CO beam sizes are indicated in the lower left corner. The white crosses indicate the locations of "Baade's blue stars," bright blue objects originally reported by Baade (1951) and studied in more detail by Lee et al. (1993) and Martínez-Delgado et al. (1999). According to the latter authors, these are probably young star clusters or associations with a minimum age of 100 Myr. The positions of these objects were taken from Table 2 of Martínez-Delgado et al. (1999).

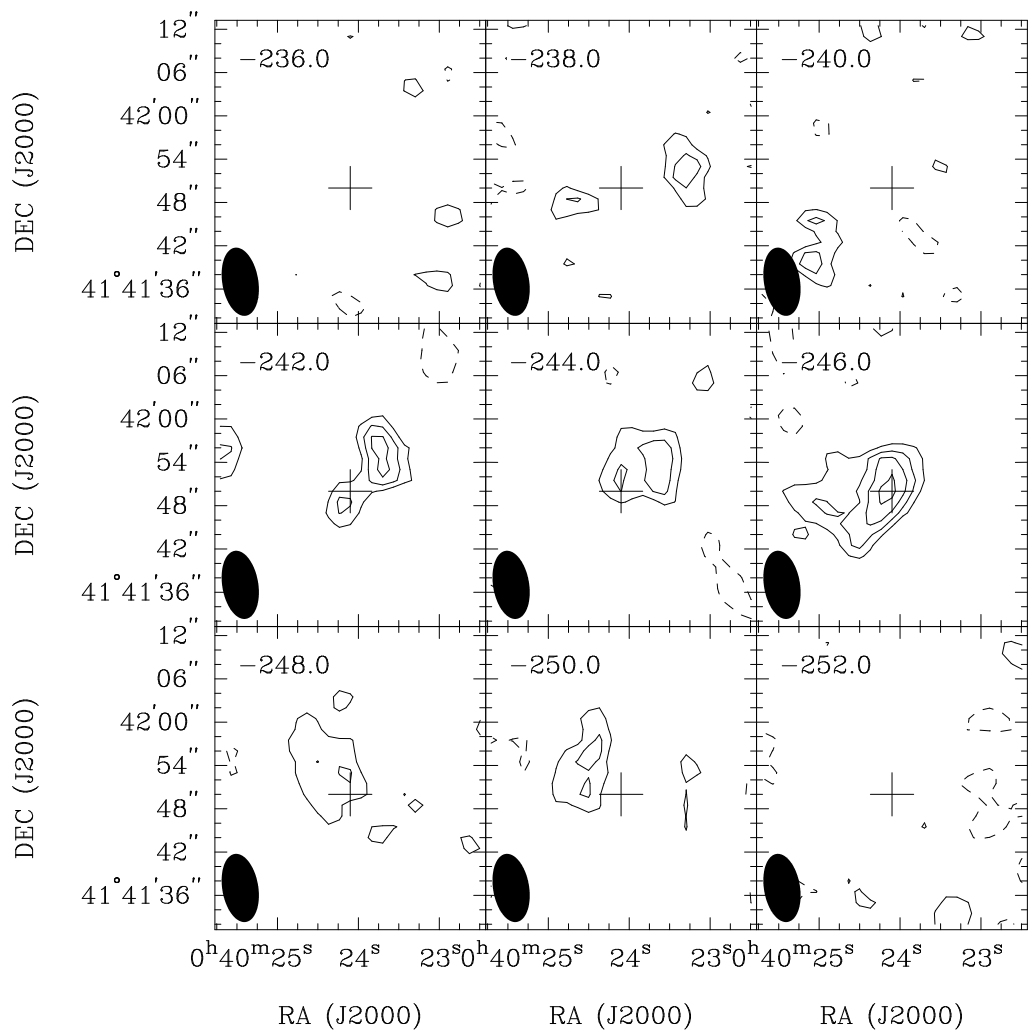

Fig. 5.— Channel maps of the CO emission in NGC 205. The contour levels are −2, 2, 3, 4 times the rms noise level (about 0.095 Jy/b = 0.18 K). Each channel is 2 km s$^{-1}$ wide and labelled with its center velocity (LSR). The beam size (9.6″× 5.0″; this is the high resolution cube) is indicated with a filled ellipse. The cross marks the pointing center of these BIMA observations and is included for a visual reference point only.

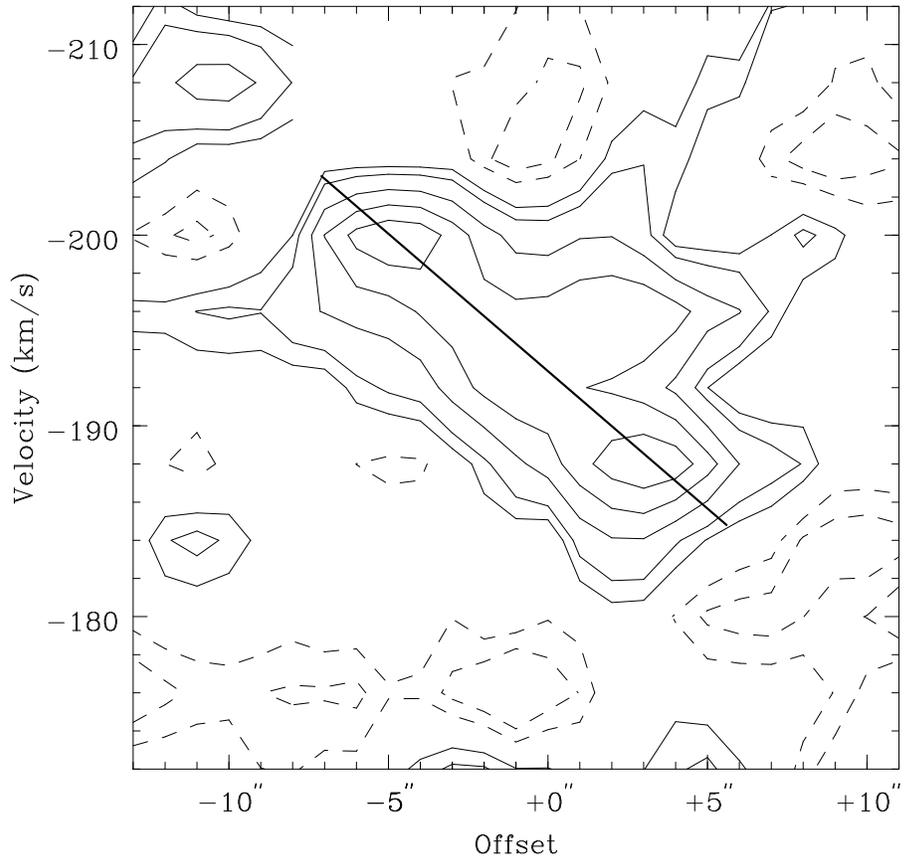

Fig. 6.— Position-velocity diagram of the molecular cloud DC1 in NGC 185. The image was made by resampling the original data cube along axes rotated at a position angle of 30° (roughly along the major axis of the cloud) and integrating parallel to the minor axis. The horizontal axis ("offset") is measured with respect to the rotation center, which has no physical significance. Contour levels are −20, −10, 10, 30, 50, 70, and 90 percent of the peak. The line segment indicates the assumed velocity gradient and the size ($2R$) and width ($2V$) of the rotating region are measured at the ends of that line.

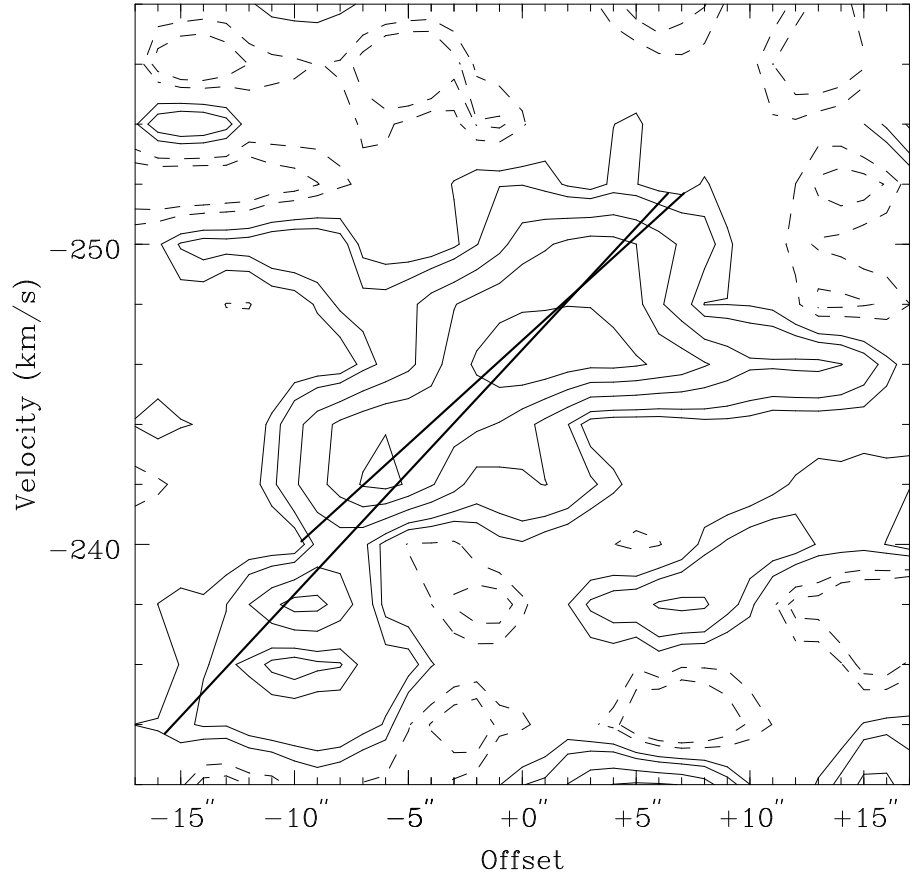

Fig. 7.— Position-velocity diagram of the molecular cloud DC11 in NGC 205. The figure was made in the same way as for NGC 185 but at a position angle of 110°. The two line segments indicate two possible gradients which differ by whether the local maximum at an offset of $-10''$ and a velocity of $-238$ km s$^{-1}$ is included or excluded. That feature can be seen as the brightest peak in the channel map at $-238$ km s$^{-1}$ (Figure 5). Another feature at $-10''$ and $-236$ km s$^{-1}$ is a local minimum.

Table 1.  Dynamical Masses and Inferred $H_2$/CO Factors

|  | NGC 185 DC1 | NGC 205 DC11 |
|---|---|---|
| Linear resolution (pc) | $17 \times 14$ | $40 \times 20$ |
| $M_{XCO}$ ($10^5$ M$_\odot$) | 0.56±0.12 | 1.16±0.36 |
| M$_{HI}$ ($10^5$ M$_\odot$) | 0.095±0.019 | 0.73±0.07 |
| Angular velocity (km s$^{-1}$ pc$^{-1}$) | 0.47±0.09 | 0.16±0.03[a] |
| $2R$ (pc) | 40±5 | 70±10 |
| $2V$ (km s$^{-1}$) | 19±2 | 11±1 |
| $M_{rot}$ ($10^5$ M$_\odot$) | 4.2±1.0 | 2.6±0.6 |
| FWHM diameters (pc) | (34±3) × (20±3) | (62±8) × (27±5) |
| D (pc)[b] | 37.5±3.5 | 62.3±8.4 |
| FWHM $\Delta v$ (km s$^{-1}$)[c] | 18.3±0.6 | 7.3±0.3 |
| $M_{vir}$ ($10^5$ M$_\odot$) | 12±1.4 | 3.3±0.52 |
| $X_{vir}/X_{gal}$ | 21±4.8[d] | 3.0±0.76[e] |
| $X_{rot}/X_{gal}$ | 7.5±2.4[d] | 2.4±0.73[e] |

[a]Measured from the shallower of the two gradients shown in Figure 7. The line with steeper gradient has angular velocity 0.20±0.03 km s$^{-1}$, velocity width ($2V$) 18±2 km s$^{-1}$, cloud diameter ($2R$) 92±13 pc, and $M_{rot}$ (8.7±2.0)×$10^5$ M$_\odot$, giving $X_{rot}/X_{gal}$~7.9±2.4.

[b]Following Wilson & Scoville (1990), $D$ is the average of the two FWHM diameters, multiplied by 1.4. The factor of 1.4 is intended to correct for the fact that the diameter used in the virial theorem should be the outer edge of the cloud, not the FWHM.

[c]The FWHM from a Gaussian fit to the $^{12}$CO J=1-0 line.

[d]The HI mass which is associated with the cloud in NGC 185 is so small that it does not appreciably affect the mass estimates or inferred conversion factors.

[e]If the HI clump is assumed to contribute to the dynamical mass estimates in NGC 205 (which is not entirely obvious, since the HI may be in a layer outside the molecular gas), the inferred conversion factors drop to $X_{vir}/X_{gal}$=2.3±0.6 and $X_{rot}/X_{gal}$=1.7±0.52.

Note. — All mass estimates include helium. Error estimates for linear sizes and masses include uncertainties in the distances of the galaxies in addition to measurement errors.